\documentclass[fleqn,usenatbib,letters]{mnras}

\usepackage{newtxtext,newtxmath}
\usepackage[T1]{fontenc}
\usepackage{ae,aecompl}

\usepackage{graphicx}	% Including figure files
\usepackage{amsmath}	% Advanced maths commands
\usepackage{amssymb}	% Extra maths symbols
\usepackage{multirow}   %Multirow for tables
\usepackage[normalem]{ulem}
\usepackage{color}
\usepackage[dvipsnames]{xcolor}
\definecolor{teal}{rgb}{0.0, 0.5, 0.5}

\title[No pulsations in VV 47]{On $\epsilon$-mechanism driven pulsations in VV 47}

\author[P. Sowicka et al.]{
Paulina Sowicka,$^{1}$\thanks{E-mail: paula@camk.edu.pl}
Gerald Handler,$^{1}$
and David Jones$^{2,3}$
\\
$^{1}$Nicolaus Copernicus Astronomical Center, Bartycka 18, PL-00-716 Warsaw, Poland\\
$^{2}$Instituto de Astrof\'isica de Canarias, E-38205 La Laguna, Tenerife, Spain\\
$^{3}$Departamento de Astrof\'isica, Universidad de La Laguna, E-38206 La Laguna, Tenerife, Spain
}

\date{Accepted XXX. Received YYY; in original form ZZZ}

\pubyear{2018}

\begin{document}
\label{firstpage}
\pagerange{\pageref{firstpage}--\pageref{lastpage}}
\maketitle
\begin{abstract}
We report new observations of the central star of the planetary nebula VV 47 carried out to verify earlier assertions that the short-period pulsation modes detected in the star are driven by the $\epsilon$ mechanism. In our data, VV 47 was not variable up to a limit of 0.52 mmag in the Fourier amplitude spectrum up to the Nyquist frequency of 21.7 mHz. Given this null result we re-analyzed the data set in which oscillations were claimed. After careful data reduction, photometry, extinction correction, and analysis with a conservative criterion of S/N $\geq$ 4 in the Fourier amplitude spectrum, we found that the star was not variable during the original observations. The oscillations reported earlier were due to an over-optimistic detection criterion. We conclude that VV 47 did not pulsate during any measurements at hand; the observational detection of $\epsilon$-driven pulsations remains arduous.
\end{abstract}


%%%%%%%%%%%%%%%%%%%%%%%%%%%%%%%%%%%%%%%%%%%%%%%%%%

%%%%%%%%%%%%%%%%% BODY OF PAPER %%%%%%%%%%%%%%%%%%

%%%%%%%%%%%%%%%%% INTRODUCTION %%%%%%%%%%%%%%%%%%%

\section{Introduction}
It has been almost 100 years since Sir Arthur Eddington, in his book ``The Internal Constitution of the Stars'' \citep{Eddington1926}, suggested the existence of a pulsational driving mechanism dependent on the nuclear energy generation rate, now known as \textit{the $\epsilon$ mechanism} (and referred to historically as ``nuclear driving''). Briefly, the $\epsilon$-mechanism driving is caused by the strong dependence of nuclear burning rates on temperature. The layers of a star where nuclear reactions take place are compressed by the enhancement of nuclear energy release causing a gain in thermal energy, while the opposite happens when these layers expand giving back the energy \citep{Unno1989}. 
The existence of such a mechanism operating in a star is somewhat obvious, although in most classes of stars it is apparently too weak to drive any pulsations at a detectable level, or at all.
For a more detailed description of the $\epsilon$ mechanism we refer the reader to, e.g., the paper by \citet{Kawaler1988}.

In the decades following Eddington's work, there was much interest in this mechanism from the community, giving as example the work by \citet{Ledoux1941} and \citet{SchwHarm1959} that suggested that this mechanism may be responsible for the pulsational instability of very massive stars. 
In the early 1980s - early 1990s, it was suggested that the instability caused by the $\epsilon$-mechanism might also operate in Wolf-Rayet stars, possibly driving their strong stellar winds. In the best candidate (WR40), \citet{Blecha1992} detected a 627-s periodicity, but were unable to confidently assign it to the $\epsilon$-mechanism. A later study by \citet{Bratschi1996} could not confirm this detection. Motivated by theoretical calculations suggesting that M dwarf stars can experience pulsational instability driven by nuclear burning, \citet{Baran2011} carried out a survey for variability among such stars, but failed to detect any variability consistent with this idea.
\citet{Miller2011} attributed the observed variability of the hot subdwarf B star LS IV-14$^{\circ}$116 to nonradial g-mode pulsations excited by the $\epsilon$-mechanism, operating in He-burning shells that appear before the star settles in the He-core burning phase. \citet{Randall2015}, disputed this conclusion based on their determinations of the stellar parameters being more consistent with a star on the Helium main sequence. In turn, \citet{battich2018} argued that these parameters may be consistent with their models of pre-horizontal branch stars, and the $\epsilon$-mechanism could still excite pulsations with periods roughly similar to those observed.

Theoretical calculations have shown that pulsations can also be driven in pre-white dwarfs via the $\epsilon$ mechanism operating in the remnants of nuclear burning in their envelopes \citep{Kawaler1986}. The mechanism operating in such conditions is not very effective, and could possibly only be responsible for driving the lowest order g-modes, which have not yet been seen in these stars. 

One of the candidates to show such pulsations is the central star of the planetary nebula (PN) VV 47 (hereinafter referred to as VV 47). Observations by \citet{Liebert1988} in 1984, found the star to be variable across multiple nights, but with the source of the variability remaining elusive. They summarized by saying that any real periodicities range from tens of minutes to several hours, and concluded: ``However, any real light variability in this object appears to be irregular in nature, and may be due to a different mechanism than the types of pulsations believed to be responsible in K~1-16, PG 1159-035, and similar stars.'' A few years later \citet{CiardulloBond1996} observed VV 47 on four nights between 1987 and 1990. They did not confirm the tentative variability reported by \citet{Liebert1988}, but neither could they rule out very low amplitude variations between nights. 

Finally, \citet{GonzalezPerez2006} (hereinafter GP06) concluded VV 47 to be a low-amplitude pulsator with an extremely complicated power spectrum based on multiple nights of observations using the 2.56-m Nordic Optical Telescope (NOT). In addition to some low-frequency variability, the authors explained the presence of some high-frequency peaks as possibly being driven by the $\epsilon$ mechanism, which would have been the first observational evidence for pulsational driving by this mechanism. 

Following the findings of GP06, VV 47 became the subject of extensive asteroseismic modelling. \citet{Corsico2009} presented a fully non-adiabatic stability analysis based on their PG~1159 models and explored the possibility of pulsational driving by the $\epsilon$ mechanism. The authors found strong evidence for the existence of an additional instability strip originating from the short-period g modes excited by the $\epsilon$ mechanism, with VV 47 lying in the region where both $\kappa$- and $\epsilon$-destabilized modes are predicted. They estimated the mass from the period spacing data, with the mean period spacing of about 24s, to be about $0.52-0.53$ M$_{\odot}$, in perfect agreement with the spectroscopic mass ($\approx$ 0.525 M$_{\odot}$) also derived by \citet{Corsico2009}. Although the uncertainties on T$_{\text{eff}}$ -- $\log g$ for VV 47 are so large, that the error box covers evolutionary tracks for masses between about 0.51--0.59 M$_{\odot}$, the two independent mass estimations being in such a good agreement narrows the possible mass range down. 
A stability analysis of all possible oscillations listed by GP06 suggested that if all reported modes were real, the shortest periodicities would be due to the $\epsilon$ mechanism. When taking into account only the modes with the highest probability of being real according to GP06, the shortest period mode over the FAP of GP06 ($\sim$261 s) would be too long to be explained as $\epsilon$-driven (see Fig. 5 of \citealt{Corsico2009}).

\citet{Calcaferro2016} attempted to find an asteroseismic model of the star with the main aim to derive its total mass. The authors estimated the mean period spacing based on GP06 data with three methods: the inverse variance, the Kolmogorov-Smirnov test, and the FT significance test (references in the original paper). When using the complete list of periods all of these methods were inconclusive. Instead, when rejecting one or more periods, the authors were able to find excellent agreement between the methods implying strong evidence for a constant period spacing of 24.2 s. This period spacing value together with the evolutionary models indicated a total mass of about 0.52 M$_{\odot}$, in perfect agreement with all previous estimations. However, even when taking different sets of periods and evolutionary stages of VV 47, the authors were unable to find an unambiguous model characterizing the star.

In the present work, we return to VV 47 with new observations using the 4.2-m William Herschel Telescope (WHT) in order to re-examine the possible existence of the $\epsilon$-driven pulsations reported by GP06. 

%%%%%%%%%%%%%%%%%%%%%%%%%%%%%%%%%%%%%%%%%%%%%%%%%%%%%%%%%%%%

%%%%%%%%%%%%%%%%%%%% NEW WHT OBSERVATIONS %%%%%%%%%%%%%%%%%%

\section{New WHT observations}
\begin{figure*}
   \centering
   \includegraphics[width=\textwidth,viewport=00 35 1090 410]{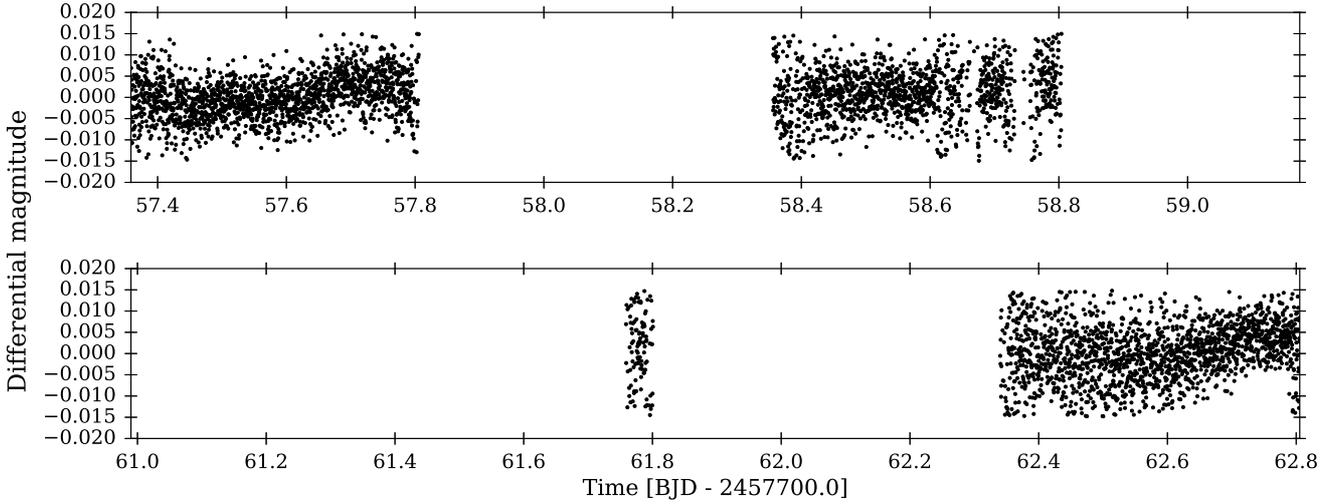}
   \label{fig:lightcurve1}
   \caption{Light curve of VV 47 from our WHT observations. Note the discontinuity between the two panels.}
\end{figure*}

VV 47 was observed using the WHT located at the Observatorio del Roque de 
los Muchachos (La Palma, Spain), equipped with ACAM, on $3-8$ January 2017. The log of observations is presented in Table~\ref{table:logWHT}. We used exposure times of about 20-s, varying it slightly from night to night to suppress aliasing at the sampling frequency. We also used windowing to reduce the readout time to about 3s. Observations were started in dark time ($3-4$ Jan) with growing influence of the Moon over the next nights, up to 84\% illumination at 60$^{\circ}$ from the target on the last night. 
The total on-target time was 33h 48m, comprising two complete nights, another which was frequently interrupted by passing clouds and a fourth night on which only one hour of observations could be acquired (see Fig. \ref{fig:lightcurve1} and Table \ref{table:logWHT}).  
The data were reduced using our own Python-based routines based mostly on \texttt{astropy} package, which consisted of standard bias subtraction, flat-fielding, and aligning the data. 

\begin{figure*}
   \centering
   \includegraphics[width=\textwidth,viewport=00 35 870 270]{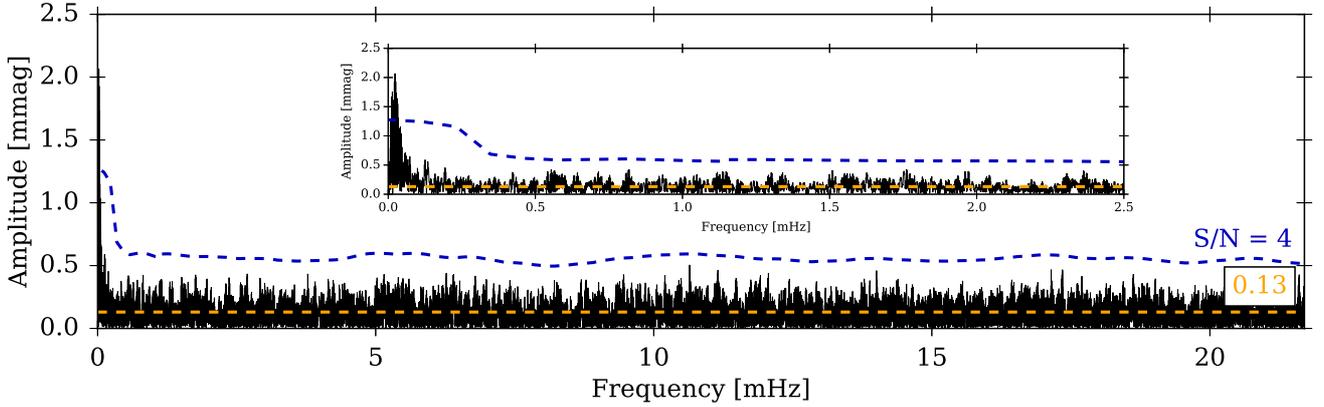}
   \label{fig:vv47fourier}
   \caption{The Fourier amplitude spectrum of all four nights of VV 47 WHT data. There are no significant peaks in the spectrum, except for a low-frequency signal close to 2 cycles per sidereal day (which is about 0.023 mHz) that we consider of instrumental origin. A zoom of the low-frequency region is presented in inset. 
}
\end{figure*} 

\begin{table*}
\begin{tabular}{c|c|c|c|c|c|c|c}
JD-2457700.0 & Exp. time (s) & N$_{i}$ & Total time & Filter & Min-median-max FWHM & Aperture scaling & rms scatter \\
& & & & & of the night (arcsec) & factor & (mmag) \\ \hline
57.3562262-57.8034369 & 20.0 & 1679 & 10h 44min & Sloan G & 0.86 - 1.46 - 4.24 & 1.37 & 5.08 \\
58.3536221-58.8025573 & 20.2 & 1671 & 10h 46min & Sloan G & 1.02 - 1.79 - 4.76 & 1.20 & 10.55\\
61.7569323-61.7997911 & 20.3 & 159 & 1h 2min & Sloan G & 3.55 - 4.97 - 7.41 & 0.85 & 15.86\\
62.3343397-62.8041082 & 19.7  & 1787 & 11h 16min  & Sloan G & 0.88 - 1.54 - 3.06  & 1.27 & 7.27\\
\label{table:logWHT}
\end{tabular}
\caption{The log of our new observations with the WHT.}
\end{table*}

%%%%%%%%%%%%%%%%%%%% Photometry %%%%%%%%%%%%%%%%%%

\subsection{Photometry}
\label{sec:photWHT}
The varying conditions during our observations made optimising the photometric measurements challenging. We found that traditional aperture photometry gave the best results when the aperture size was scaled with seeing. We used a trial and error approach to find the optimal scaling factor for the aperture size for a given night's observations by minimizing the rms scatter (Table~\ref{table:logWHT}) of photometric measurements extracted using the Starlink \texttt{autophotom} package \citep{photomascl}. In our windowed field of view, there were three other stars bright enough to be used as possible comparison stars. We achieved the best results in terms of signal-to-noise ratio (SNR) when using an `artificial' comparison star comprising the summed flux from all three available comparison stars. Because our target is much hotter than the comparison stars it was necessary to correct the resulting light curves for differential colour extinction, which was done using the first approximation linear fit to the $\Delta$mag($\sec z$) relation. The last step consisted of rejecting outlying points with a dedicated program that runs a moving-average filter over the data and rejects the most obvious single outliers. The final light curve is shown in Fig.~\ref{fig:lightcurve1}. The time scales of the remaining, low-amplitude long-term trends are too long to be due to stellar pulsation. 

%%%%%%%%%%%%%%%%%%%% Pulsations of VV 47? %%%%%%%%%%%%%%%%%%

\subsection{Pulsations of VV 47?}
We calculated the Fourier transform (FT) amplitude spectrum of all nights of VV 47 data (Fig.~\ref{fig:vv47fourier}) up to the Nyquist frequency of the data set ($\approx 21.7$\,mHz) using \texttt{Period04} \citep{Period04}. Aside from the manifestation of the above mentioned long-term trend, the spectrum is flat with no significant peaks up to a limit of 0.52 mmag.
This detection limit was calculated following the conservative criterion that for any peak to be reliably detected it must fulfill S/N$\geq$4 in amplitude \citep{Breger1993}. We then calculated the amplitude FT for each night separately, again not finding any statistically significant peaks. With this null result we conclude that VV 47 did not pulsate during our observations, and hence might not be a pulsator at all - and perhaps not even a variable star.

%%%%%%%%%%%%%%%%%%%%%%%%%%%%%%%%%%%%%%%%%%%%%%%%%%%%%%%%%%%%%%%%

%%%%%%%%%%%%%%%%%%%% RE-ANALYSIS OF GP06 DATA %%%%%%%%%%%%%%%%%%

\section{Re-analysis of archival data}
Intrigued by the apparent absence of not only the pulsation modes detected by previous authors, but any pulsations at all, we re-examined the data set of GP06. These authors reported the results of a survey for photometric variability among 11 hot, H-deficient PN nuclei (PNNi), carried out in years 2000-2001 with the NOT. VV 47 was observed on three nights for a total of 22.2 h using the ALFOSC instrument; for details of these observations see GP06.

We retrieved the data of GP06 from the NOT archive which did not include the final light curves their analysis was based upon. Therefore we decided to reduce the original CCD frames anew. To this end, the original frames were converted to 2D images using the \texttt{rtconv} program \citep{Ostensen2000PhD}, which was adapted to run on modern computers (G. Stachowski, private communication). Bias and flat field, but no gain correction were applied. Next we performed photometry, using the same approach as described in Section~\ref{sec:photWHT}. The conditions during each run varied but ultimately the best aperture scaling factor was found to be 1.3 for all runs. A differential light curve was created dividing the flux from the target by the summed flux of two comparison stars available during each run. We corrected for differential colour extinction following the methodology described in Section~\ref{sec:photWHT}, with the final estimated rms scatter per data point in each run being about $4-6$ mmag (Table~\ref{table:logGP}). Next, we calculated amplitude FTs for each run separately up to the Nyquist frequency, i.e. runs 1 - 3: $\approx$ 720 c/d = 8333 $\mu$Hz, run 4: 1080 c/d = 12500 $\mu$Hz. 

\begin{table*}
\begin{tabular}{c|c|c|c|c|c|c|c|c}
Run&UT start date & Duty cycle (s) & N$_{i}$ & Total time & Filter & Min-median-max FWHM & Aperture scaling & rms scatter \\
& & & & & & of the night (arcsec) & factor & (mmag) \\ \hline
1 & 2001-01-16T00:36:53 & 60.0 & 348 & 5h47min & B & 1.09 - 1.33 - 2.18 & 1.3 & 4.377 \\
2 & 2001-01-16T21:04:03 & 60.0 & 377 & 6h16min & B & 0.89 - 1.18 - 1.93 & 1.3 & 4.641\\
3 & 2001-01-17T05:04:47 & 60.0 & 106 & 1h45min & B & 1.20 - 1.33 - 1.87 & 1.3 & 5.351\\
4 & 2001-01-18T20:17:01 & 40.0  & 709 & 8h24min  & B & 0.91 - 1.20 - 2.18  & 1.3 & 5.541\\
\label{table:logGP}
\end{tabular}
\caption{The log of observations by GP06.}
\end{table*}

Before we proceed to the direct comparison of the results, we have to discuss two fundamental differences in our analyses compared to GP06. These authors decided to use a different approach to assess the significance of an individual peak - not our conservative criterion of S/N$\geq$4 \citep{Breger1993}, but a variant of the method described by \citet{Kepler1993}. In brief, this method assesses the probability for a given peak in the FT of being real by using the equation
\begin{equation}
P_{\text{obs}} = \ln \frac{N_i}{\text{FALSE}} \langle P \rangle,
\end{equation}
where $P_{\text{obs}}$ is the power that satisfies this criterion, ${N_i}$ - the number of independent frequencies, $\text{FALSE}$ - a false alarm probability (FAP), and $\langle P \rangle$ is the average power in the region around the frequency of interest. A FAP of 1/100 means that a peak is taken as being real only if it has less than 1 in 100 possibility to be due to noise. 

There are however differences between the original paper by \citet{Kepler1993} and the approach by GP06. $N_i$ is defined by \citet{Kepler1993} as the number of independent frequencies, while in GP06 as the number of points in the light curve. This is only correct  in the case of equally spaced data without gaps when the periodogram is computed from zero to the sampling frequency (=2$ f_{\text{Nyquist}}$). The first condition seems fulfilled by the authors as they do not mention an outlier removal procedure, plus the number of data points in their table corresponds to the number of frames taken. 
However, the second condition may not have been fulfilled - judging purely from the FTs presented in the paper - as the authors plot all FTs for VV 47 starting from 130 $\mu$Hz up to about 8500 $\mu$Hz except for the last run, where the curve exceeds 8500 $\mu$Hz but it is unclear at which value it ends. As such, this appears to be a calculation up to the Nyquist frequency rather than to the sampling frequency (i.e. 2$f_{\text{Nyquist}}$). Also, the mean power $\langle P \rangle$ would then need to be evaluated over the whole FT, which does not appear to be the case in their analysis. If this is indeed true, then the significance criterion of GP06 was incorrectly applied. Also, their choice of the FALSE value of 1/20 is a rather relaxed criterion for this kind of star, with conservative values of $\sim$ 1/1000 being more typical \citep[e.g.][]{Castanheira2013}.

GP06 justified the need of such an alteration with the statement that ``periods that have been observed for PNNi have quite low amplitude (a few mmag) and show temporal variability in the amplitude on different time scales.'' The peaks the authors found in the FTs of their targets in that way had low amplitudes and were variable on time scales of days or hours. Low amplitudes are common for this type of star, but the only known mechanisms that would make signals vary in amplitude on such short time scales would be beating of unresolved pulsation frequencies (which should become resolved and detected when analyzing longer strings of data), interference with noise peaks, or those peaks indeed being due to noise themselves. We decided to not use GP06's approach and instead remain with the classical conservative criterion of S/N$\geq$4 in amplitude \citep{Breger1993}.

\begin{figure}
   \centering
   \includegraphics[width=0.5\textwidth,viewport=00 45 470 285]{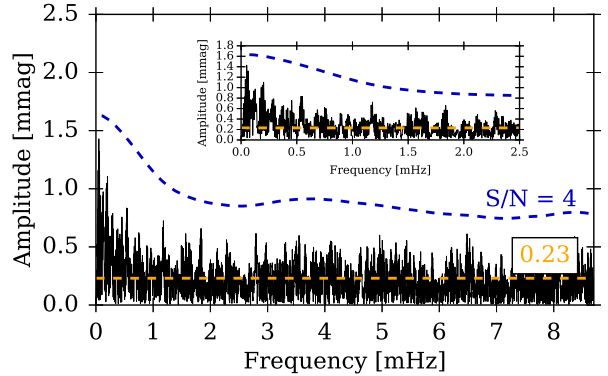}
   \label{fig:vv47fourierGP}
   \caption{The Fourier amplitude spectrum of \citet{GonzalezPerez2006} data after our reduction.
   }
\end{figure} 

In their first run (15/16.01.2001) GP06 estimated the average power to be 0.15 $\mu$mp, while our estimation was 0.22 $\mu$mp. The average power estimated by GP06 appears over-optimistic (their Figs. $8-9$). They detected four peaks, with two of them above their FAP. We did not detect any significant peaks above our criterion (no peaks were even close to our significance curve).

For the second run (16/17.01.2001) the average power was estimated by GP06 to be 0.17 $\mu$mp, while we estimated it to be 0.23 $\mu$mp. Three peaks satisfied their criterion with two of them being strong candidates for real ones - we found no peaks satisfying our criterion.

In the third run (16/17.01.2001) GP06 estimated the average power to be 1.03 $\mu$mp, while our estimate was 1.08 $\mu$mp, close to their value. The one order of magnitude higher noise level for this run is due to its very short duration (less than 2 hours). GP06 found one peak satisfying the criterion, again we found none.

For the final run the authors estimated the average power to be 0.42 $\mu$mp, while our value was 0.17 $\mu$mp. This time the value obtained by GP06 is suspiciously high. This run was the longest and, as such, the noise level should be the lowest. They claimed two peaks above the FAP, while we did not find any. The most interesting result of GP06 was the possible detection of the highest frequency peak attributed to the $\epsilon$ mechanism in the first section of the run lasting 2.3 hr. We scrutinized this section of the run in our data and found no evidence for the claimed oscillations.

In Fig.~\ref{fig:vv47fourierGP}, we present our overall FT of the GP06 data. There are no significant peaks up to a 0.92 mmag level. This level is almost two times higher than that of our new observations, highlighting a significant reduction in the detection limit of the new observations. No variability found at such a low level, neither in the range where $\epsilon$-driven oscillations, nor where $\kappa$-driven modes are expected (insets in Figs.~\ref{fig:vv47fourier} and \ref{fig:vv47fourierGP}), suggests that VV 47 is not even a variable star within the accuracy of our measurements.

%%%%%%%%%%%%%%%%%%%%%%%%%%%%%%%%%%%%%%%%%%%%%%%%%%%%%%%%%%%%%%%

%%%%%%%%%%%%%%%%%%%% SUMMARY AND CONCLUSIONS %%%%%%%%%%%%%%%%%%

\section{Summary and conclusions}

We have presented a variability study of VV 47 based on new observations as well as a re-analysis of previously published data (GP06). Our new observations, made using the 4.2-m WHT, arrive at a significantly lower detection limit than the previously published data. Adopting the classical significance criterion of S/N$\geq$4 in the Fourier amplitude spectrum, the detection limit was 0.52 mmag for the new data, and 0.92 mmag for the archival data. Even at such a low limit, we were unable to detect significant peaks in either data set, neither in the high-frequency domain where $\epsilon$-driven pulsations may be present, nor at lower frequencies where possible $\kappa$-driven oscillations are expected. Instead, we attribute GP06's claims for pulsations in VV 47 to a particularly relaxed significance criterion for the detection of periodicities in their data set (especially for stars of this type). In conclusion, no pulsations or demonstrably intrinsic variability were detected during either observing campaign. Therefore VV 47 is not only a non-pulsator; it may not even be a variable star. 

In any case, the findings of GP06 were exciting enough that theoretical work on the star soon followed. \citet{Corsico2009} took the highest-frequency peaks as candidates for low-$k$-order $g$-modes excited by the $\epsilon$ mechanism and estimated the mass from the period spacing data. Interestingly, the noise peaks detected by GP06 spaced by about $20-30$ s, arrived at mass estimation which was in perfect agreement with the spectroscopic mass of VV 47. Given the T$_{\text{eff}}$ -- $\log g$ uncertainties, it is not possible to find an unambiguous model without further constraints, which in this case were overinterpreted asteroseismic measurements. The true mass of VV 47 hence might be different, and what follows, the evolutionary tracks used for the model may not be appropriate since these tracks characterize e.g. the thickness of the He-rich envelope, which is crucial for the assessment of the efficiency of the $\epsilon$ mechanism linked to the active He-burning shell. With the evolutionary sequence of M$_*$ = 0.530 M$_{\odot}$ and all periods reported by GP06 the authors could attribute most of the longest period modes to the $\kappa$ mechanism and the shortest ones to the $\epsilon$ mechanism. However, when taking into account only the modes described by GP06 as having the best chances of being real, the shortest-period mode would not be explicable as destabilized by the $\epsilon$ mechanism. This can be taken as an argument that the physically sound models of \citet{Corsico2009} were not in agreement with the periodicities reported by GP06.

Interestingly, the spurious periodicities of GP06 led \citet{Calcaferro2016} into finding a period spacing value for VV 47 in excellent agreement within three used methods, with a mass estimation based on this spacing agreeing with all previous determinations. Even though the estimated parameters looked correct, \citet{Calcaferro2016} were not able to find an unambiguous model for VV 47.

The example of VV 47 shows that it is possible to derive credible model fits even if based on inadequate data, in addition even being in agreement with values determined using other methods. Careful analysis and interpretation of observational data should therefore prevail over the temptation to claim potentially exciting results on a poor base.

%%%%%%%%%%%%%%%%%%%%%%%%%%%%%%%%%%%%%%%%%%%%%%%%%%%%%%%%

%%%%%%%%%%%%%%%%%%%% ACKNOWLEDGEMENTS %%%%%%%%%%%%%%%%%%

\section*{Acknowledgements}
We thank Peter Meldgaard Sorensen and Sergio Armas Perez for their help with retrieving archival NOT data, Greg Stachowski for his help with \texttt{rtconv}, as well as Hiromoto Shibahashi, Alosza Pamyatnykh and Alejandro C\'orsico for helpful suggestions.
PS and GH thank the Polish National Center for Science (NCN) for support through grant 2015/18/A/ST9/00578. PS also acknowledges CAMK grant for young researchers. This research has been supported by the Spanish Ministry of Economy and Competitiveness (MINECO) under the grant AYA2017-83383-P.

%%%%%%%%%%%%%%%%%%%%%%%%%%%%%%%%%%%%%%%%%%%%%%%%%%


\bibliographystyle{mnras}
\bibliography{literature} % if your bibtex file is called example.bib

\begin{thebibliography}{}
\makeatletter
\relax
\def\mn@urlcharsother{\let\do\@makeother \do\$\do\&\do\#\do\^\do\_\do\%\do\~}
\def\mn@doi{\begingroup\mn@urlcharsother \@ifnextchar [ {\mn@doi@}
  {\mn@doi@[]}}
\def\mn@doi@[#1]#2{\def\@tempa{#1}\ifx\@tempa\@empty \href
  {http://dx.doi.org/#2} {doi:#2}\else \href {http://dx.doi.org/#2} {#1}\fi
  \endgroup}
\def\mn@eprint#1#2{\mn@eprint@#1:#2::\@nil}
\def\mn@eprint@arXiv#1{\href {http://arxiv.org/abs/#1} {{\tt arXiv:#1}}}
\def\mn@eprint@dblp#1{\href {http://dblp.uni-trier.de/rec/bibtex/#1.xml}
  {dblp:#1}}
\def\mn@eprint@#1:#2:#3:#4\@nil{\def\@tempa {#1}\def\@tempb {#2}\def\@tempc
  {#3}\ifx \@tempc \@empty \let \@tempc \@tempb \let \@tempb \@tempa \fi \ifx
  \@tempb \@empty \def\@tempb {arXiv}\fi \@ifundefined
  {mn@eprint@\@tempb}{\@tempb:\@tempc}{\expandafter \expandafter \csname
  mn@eprint@\@tempb\endcsname \expandafter{\@tempc}}}

\bibitem[\protect\citeauthoryear{{Baran} et~al.,}{{Baran}
  et~al.}{2011}]{Baran2011}
{Baran} A.~S.,  et~al., 2011, \actaa, \href
  {http://adsabs.harvard.edu/abs/2011AcA....61...37B} {61, 37}

\bibitem[\protect\citeauthoryear{{Battich}, {Miller Bertolami}, {C{\'o}rsico}
  \& {Althaus}}{{Battich} et~al.}{2018}]{battich2018}
{Battich} T.,  {Miller Bertolami} M.~M.,  {C{\'o}rsico} A.~H.,   {Althaus}
  L.~G.,  2018, preprint, \href
  {http://adsabs.harvard.edu/abs/2018arXiv180107287B} {} (\mn@eprint {arXiv}
  {1801.07287})

\bibitem[\protect\citeauthoryear{{Blecha}, {Schaller}  \& {Maeder}}{{Blecha}
  et~al.}{1992}]{Blecha1992}
{Blecha} A.,  {Schaller} G.,   {Maeder} A.,  1992, \mn@doi [\nat]
  {10.1038/360320a0}, \href {http://adsabs.harvard.edu/abs/1992Natur.360..320B}
  {360, 320}

\bibitem[\protect\citeauthoryear{{Bratschi} \& {Blecha}}{{Bratschi} \&
  {Blecha}}{1996}]{Bratschi1996}
{Bratschi} P.,  {Blecha} A.,  1996, \aap, \href
  {http://adsabs.harvard.edu/abs/1996A%26A...313..537B} {313, 537}

\bibitem[\protect\citeauthoryear{{Breger} et~al.,}{{Breger}
  et~al.}{1993}]{Breger1993}
{Breger} M.,  et~al., 1993, \aap, \href
  {http://adsabs.harvard.edu/abs/1993A%26A...271..482B} {271, 482}

\bibitem[\protect\citeauthoryear{{Calcaferro}, {C{\'o}rsico}  \&
  {Althaus}}{{Calcaferro} et~al.}{2016}]{Calcaferro2016}
{Calcaferro} L.~M.,  {C{\'o}rsico} A.~H.,   {Althaus} L.~G.,  2016, \mn@doi
  [\aap] {10.1051/0004-6361/201527996}, \href
  {http://adsabs.harvard.edu/abs/2016A%26A...589A..40C} {589, A40}

\bibitem[\protect\citeauthoryear{{Castanheira}, {Kepler}, {Kleinman}, {Nitta}
  \& {Fraga}}{{Castanheira} et~al.}{2013}]{Castanheira2013}
{Castanheira} B.~G.,  {Kepler} S.~O.,  {Kleinman} S.~J.,  {Nitta} A.,   {Fraga}
  L.,  2013, \mn@doi [\mnras] {10.1093/mnras/sts474}, \href
  {http://adsabs.harvard.edu/abs/2013MNRAS.430...50C} {430, 50}

\bibitem[\protect\citeauthoryear{{Ciardullo} \& {Bond}}{{Ciardullo} \&
  {Bond}}{1996}]{CiardulloBond1996}
{Ciardullo} R.,  {Bond} H.~E.,  1996, \mn@doi [\aj] {10.1086/117967}, \href
  {http://adsabs.harvard.edu/abs/1996AJ....111.2332C} {111, 2332}

\bibitem[\protect\citeauthoryear{{C{\'o}rsico}, {Althaus}, {Miller Bertolami},
  {Gonz{\'a}lez P{\'e}rez}  \& {Kepler}}{{C{\'o}rsico}
  et~al.}{2009}]{Corsico2009}
{C{\'o}rsico} A.~H.,  {Althaus} L.~G.,  {Miller Bertolami} M.~M.,
  {Gonz{\'a}lez P{\'e}rez} J.~M.,   {Kepler} S.~O.,  2009, \mn@doi [\apj]
  {10.1088/0004-637X/701/2/1008}, \href
  {http://adsabs.harvard.edu/abs/2009ApJ...701.1008C} {701, 1008}

\bibitem[\protect\citeauthoryear{{Eaton}, {Draper}, {Allan}, {Naylor}, {Mukai},
  {Currie}  \& {McCaughrean}}{{Eaton} et~al.}{2014}]{photomascl}
{Eaton} N.,  {Draper} P.~W.,  {Allan} A.,  {Naylor} T.,  {Mukai} K.,  {Currie}
  M.~J.,   {McCaughrean} M.,  2014, {PHOTOM: Photometry of digitized images},
  Astrophysics Source Code Library (\mn@eprint {ascl} {1405.013})

\bibitem[\protect\citeauthoryear{{Eddington}}{{Eddington}}{1926}]{Eddington1926}
{Eddington} A.~S.,  1926, {The Internal Constitution of the Stars}

\bibitem[\protect\citeauthoryear{{Gonz{\'a}lez P{\'e}rez}, {Solheim}  \&
  {Kamben}}{{Gonz{\'a}lez P{\'e}rez} et~al.}{2006}]{GonzalezPerez2006}
{Gonz{\'a}lez P{\'e}rez} J.~M.,  {Solheim} J.-E.,   {Kamben} R.,  2006, \mn@doi
  [\aap] {10.1051/0004-6361:20053468}, \href
  {http://adsabs.harvard.edu/abs/2006A%26A...454..527G} {454, 527}

\bibitem[\protect\citeauthoryear{{Kawaler}}{{Kawaler}}{1988}]{Kawaler1988}
{Kawaler} S.~D.,  1988, \mn@doi [\apj] {10.1086/166832}, \href
  {http://adsabs.harvard.edu/abs/1988ApJ...334..220K} {334, 220}

\bibitem[\protect\citeauthoryear{{Kawaler}, {Winget}, {Hansen}  \&
  {Iben}}{{Kawaler} et~al.}{1986}]{Kawaler1986}
{Kawaler} S.~D.,  {Winget} D.~E.,  {Hansen} C.~J.,   {Iben} Jr. I.,  1986,
  \mn@doi [\apjl] {10.1086/184701}, \href
  {http://adsabs.harvard.edu/abs/1986ApJ...306L..41K} {306, L41}

\bibitem[\protect\citeauthoryear{{Kepler}}{{Kepler}}{1993}]{Kepler1993}
{Kepler} S.~O.,  1993, \mn@doi [Baltic Astronomy] {10.1515/astro-1993-3-425},
  \href {http://adsabs.harvard.edu/abs/1993BaltA...2..515K} {2, 515}

\bibitem[\protect\citeauthoryear{{Ledoux}}{{Ledoux}}{1941}]{Ledoux1941}
{Ledoux} P.,  1941, \mn@doi [\apj] {10.1086/144359}, \href
  {http://adsabs.harvard.edu/abs/1941ApJ....94..537L} {94, 537}

\bibitem[\protect\citeauthoryear{{Lenz} \& {Breger}}{{Lenz} \&
  {Breger}}{2005}]{Period04}
{Lenz} P.,  {Breger} M.,  2005, \mn@doi [Communications in Asteroseismology]
  {10.1553/cia146s53}, \href
  {http://adsabs.harvard.edu/abs/2005CoAst.146...53L} {146, 53}

\bibitem[\protect\citeauthoryear{{Liebert}, {Fleming}, {Green}  \&
  {Grauer}}{{Liebert} et~al.}{1988}]{Liebert1988}
{Liebert} J.,  {Fleming} T.~A.,  {Green} R.~F.,   {Grauer} A.~D.,  1988,
  \mn@doi [\pasp] {10.1086/132154}, \href
  {http://adsabs.harvard.edu/abs/1988PASP..100..187L} {100, 187}

\bibitem[\protect\citeauthoryear{{Miller Bertolami}, {C{\'o}rsico}  \&
  {Althaus}}{{Miller Bertolami} et~al.}{2011}]{Miller2011}
{Miller Bertolami} M.~M.,  {C{\'o}rsico} A.~H.,   {Althaus} L.~G.,  2011,
  \mn@doi [\apjl] {10.1088/2041-8205/741/1/L3}, \href
  {http://adsabs.harvard.edu/abs/2011ApJ...741L...3M} {741, L3}

\bibitem[\protect\citeauthoryear{{{\O}stensen}}{{{\O}stensen}}{2000}]{Ostensen2000PhD}
{{\O}stensen} R.,  2000, {Time Resolved CCD Photometry}

\bibitem[\protect\citeauthoryear{{Randall}, {Bagnulo}, {Ziegerer}, {Geier}  \&
  {Fontaine}}{{Randall} et~al.}{2015}]{Randall2015}
{Randall} S.~K.,  {Bagnulo} S.,  {Ziegerer} E.,  {Geier} S.,   {Fontaine} G.,
  2015, \mn@doi [\aap] {10.1051/0004-6361/201425251}, \href
  {http://adsabs.harvard.edu/abs/2015A%26A...576A..65R} {576, A65}

\bibitem[\protect\citeauthoryear{{Schwarzschild} \& {H{\"a}rm}}{{Schwarzschild}
  \& {H{\"a}rm}}{1959}]{SchwHarm1959}
{Schwarzschild} M.,  {H{\"a}rm} R.,  1959, \mn@doi [\apj] {10.1086/146662},
  \href {http://adsabs.harvard.edu/abs/1959ApJ...129..637S} {129, 637}

\bibitem[\protect\citeauthoryear{{Unno}, {Osaki}, {Ando}, {Saio}  \&
  {Shibahashi}}{{Unno} et~al.}{1989}]{Unno1989}
{Unno} W.,  {Osaki} Y.,  {Ando} H.,  {Saio} H.,   {Shibahashi} H.,  1989,
  {Nonradial oscillations of stars}

\makeatother
\end{thebibliography}

%\begin{thebibliography}{99}
%\bibitem[\protect\citeauthoryear{Eddington}{1926}]{Eddington1926}
%Eddington A.~S., 1926, The Internal Constitution of the Stars, Cambridge: Cambridge University Press
%\end{thebibliography}

% Don't change these lines
\bsp	% typesetting comment
\label{lastpage}
\end{document}